\documentclass[prd,preprint,superscriptaddress,showpacs,nofootinbib,%
tightenlines ]{revtex4}
\usepackage{epsfig}
\usepackage{slashed}
\usepackage{amssymb}

\newcommand{\ben}{\begin{displaymath}}
\newcommand{\een}{\end{displaymath}}
\newcommand{\be}{\begin{equation}}
\newcommand{\ee}{\end{equation}}
\newcommand{\bea}{\begin{eqnarray}}
\newcommand{\eea}{\end{eqnarray}}
\begin{document}
\preprint{MKPH-T-06-08}
\title{Probing the convergence of perturbative series in baryon \\ chiral perturbation theory}
\author{D.~Djukanovic}
\affiliation{Institut f\"ur Kernphysik, Johannes
Gutenberg-Universit\"at, D-55099 Mainz, Germany}
\author{J.~Gegelia}
\affiliation{Institut f\"ur Kernphysik, Johannes
Gutenberg-Universit\"at, D-55099 Mainz, Germany} \affiliation{High
Energy Physics Institute, Tbilisi State University, Tbilisi,
Georgia}
\author{S.~Scherer}
\affiliation{Institut f\"ur Kernphysik, Johannes
Gutenberg-Universit\"at, D-55099 Mainz, Germany}
\date{April 19, 2006}
\begin{abstract}
Using the examples of pion-nucleon scattering and the nucleon mass
we analyze the convergence of perturbative series in the framework
of baryon chiral perturbation theory. For both cases we sum up
sets of an infinite number of diagrams by solving equations
exactly and compare the solutions with the perturbative
contributions.
\end{abstract}

\maketitle

\section{Introduction}

Nowadays, mesonic chiral perturbation theory (ChPT)
\cite{Weinberg:1979kz,Gasser:1984gg,Gasser:1984yg} is widely
accepted as the low-energy theory of the strong interactions based
on the underlying symmetries of QCD. Impressive accuracy in the
description of data has been achieved in the last decade
\cite{Colangelo:2000jc,Colangelo:2001df,Caprini:2003ta,Caprini:2005an,
Caprini:2005zr,Colangelo:2005cq,Scherer:2005ki,Bijnens:2006zp}. On
the other hand, the baryon sector of ChPT proved to be more
problematic \cite{Gasser:1988rb}. Although there has been
considerable progress in recent years (see, e.g.,
\cite{Scherer:2002tk}), the issue of the convergence of
perturbative calculations in the nucleon sector of the effective
theory is still of great interest.

To compare lattice calculations with experimental data, one has to
extrapolate the results to physical quark masses. The preferred
method to cope with this problem is to calculate the physical
quantities as functions of the quark masses in an effective
field-theoretical approach (see, e.g.,
\cite{Beane:2004ks,Procura:2003ig,Leinweber:2005cm,Procura:2006bj}).
Of course, these effective theories have a limited range of
applicability.
    The aim of this work is to probe the issue of convergence of perturbative
calculations in BChPT. To that end we consider the contribution of
particular infinite sets of diagrams to the $\pi N$ scattering
amplitude and to the nucleon self-energy. These contributions are
summed up by solving integral equations analytically. For a
related analysis of exact solutions to the Bethe-Salpeter
equations in BChPT in the $SU(3)$ sector, see
Ref.~\cite{Lutz:2001yb}.

Throughout this paper we use dimensional regularization in
combination with the infrared renormalization scheme
\cite{Becher:1999he,Schindler:2003xv} without explicitly showing
the counter-terms responsible for the subtractions of loop
diagrams.

\section{Pion-nucleon scattering }

    First, let us specify the Lagrangians required for the purposes of
this work. From the mesonic sector we need the lowest-order
Lagrangian of the $SU(2)$ sector \cite{Gasser:1984yg}
\begin{equation}
\label{l2} {\cal L}_2=\frac{F^2}{4}\mbox{Tr}(\partial_\mu U
\partial^\mu U^\dagger) +\frac{F^2 M^2}{4}\mbox{Tr}(U^\dagger+ U),
\end{equation}
where $U$ is a unimodular unitary matrix containing the Goldstone
boson fields.
   In Eq.\ (\ref{l2}), $F$ denotes the pion-decay constant in the chiral
limit: $F_\pi=F[1+{\cal O}(\hat{m})]=92.4$ MeV.
   Here, we work in the isospin-symmetric limit $m_u=m_d=\hat{m}$,
and the lowest-order expression for the squared pion mass is
$M^2=2 B \hat{m}$, where $B$ is related to the quark condensate
$\langle \bar{q} q\rangle_0$ in the chiral limit
\cite{Gasser:1984yg}.
   Next, the nucleon fields are collected in an isospin doublet
\begin{displaymath}
\Psi=\left(\begin{array}{c}p\\n\end{array}\right)
\end{displaymath}
with two four-component Dirac fields $p$ and $n$ describing the
proton and neutron, respectively.
    From the nucleon sector we need the leading-order Lagrangian
omitting external sources
\begin{equation}
{\cal L}_{\pi N}^{(1)}=\bar \Psi \left( i\gamma_\mu D^\mu -m
+\frac{1}{2} \stackrel{\circ}{g_A}\gamma_\mu \gamma_5 u^\mu\right)
\Psi, \label{lolagr}
\end{equation}
where
$$
D_\mu\Psi = (\partial_\mu +\Gamma_\mu)\Psi,\quad u^2=U,\quad u_\mu
=iu^{\dagger}\partial_\mu U u^{\dagger},\quad \Gamma_\mu
=\frac{1}{2} [u^{\dagger},\partial_\mu u],
$$
and  $m$ and $\stackrel{\circ}{g_A}$ refer to the chiral limit of
the physical nucleon mass and the axial-vector coupling constant.

\medskip

Let us consider elastic $\pi N$ scattering with $p_1$, $q_1$ the
four-momenta of the incoming and $p_2$, $q_2$ the four-momenta of
the outgoing nucleons and pions, respectively (see Fig.~1). The
corresponding vertex function (amputated Green's function) can be
obtained by solving the integral equation
\begin{eqnarray}
\Gamma^{ba}\left( p_2,q_2;p_1,q_1\right) &=& V^{ba}\left(
p_2,q_2;p_1,q_1\right)\nonumber\\
&+& \int\frac{d^n k}{(2\,\pi )^n}\,V^{bc}\left(
p_2,q_2;p-k,k\right)\,G(p-k,k)\, \Gamma^{ca}\left(
p-k,k;p_1,q_1\right) \,, \label{Tamplequation}
\end{eqnarray}
where $p=p_1+q_1$. $V^{ba}$ stands for the $\pi N$ effective
potential and $G(p-k,k)$ is the product of the (dressed) nucleon
and pion propagators. Here, the effective potential is defined as
the sum of all diagrams contributing to the vertex function, which
cannot be reduced to two $\pi N$ scattering diagrams by cutting
one pion line and one nucleon line.

The standard representation for the vertex function in terms of
isospin symmetric and antisymmetric parts reads
\begin{equation}
\Gamma^{ba} = \delta^{ba}\,\Gamma^+ + \frac{1}{2}\,\left[
\tau^b,\tau^a\right]\, \Gamma^- \,. \label{Tampl}
\end{equation}
\noindent For our purposes it is convenient to decompose the
scattering amplitude in isospin-invariant components
\begin{eqnarray}
\Gamma^{3/2} &=& \Gamma^+-\Gamma^- \,,
\nonumber\\
\Gamma^{1/2} &=& \Gamma^+ + 2\, \Gamma^- \,.
\label{t1t2definition}
\end{eqnarray}
These vertex functions satisfy the integral equations written
symbolically as
\begin{equation}
\Gamma^{I} = V^{I} + V^{I}\,G\, \Gamma^{I} \,,
\label{decoupledequations}
\end{equation}
where $I=1/2$ or $3/2$\footnote{Everywhere below the index $I$ can
take one of the two values $1/2$ and $3/2$.} and
\begin{eqnarray}
V^{3/2} &=& V^+-V^- \,,
\nonumber\\
V^{1/2} &=& V^+ + 2\, V^- \,. \label{v1t2definition}
\nonumber\\
 V^{ba} &=& \delta^{ba}\,V^+ +
\frac{1}{2}\,\left[ \tau^b,\tau^a\right]\, V^-\,.
\end{eqnarray}

\noindent Suppose the potential can be written as
\begin{equation}
V^{I}\left( p_2,q_2;p_1,q_1\right)=\left(%
\begin{array}{cc}
  1 & q_2
\hspace{-.85em}/\hspace{.1em} \\
\end{array}%
\right)\left(%
\begin{array}{cc}
v_{11}^{I} & v_{12}^{I} \\
  v_{21}^{I} & v_{22}^{I} \\
\end{array}%
\right)\left(%
\begin{array}{c}
  1 \\
  q_1
\hspace{-.85em}/\hspace{.1em} \\
\end{array}%
\right)\,, \label{vmatrix}
\end{equation}
where the $v^{I}_{ij}$ depend only on $p=p_1+q_1=p_2+q_2$ as is
the case, e.g., for the
potential\footnote{Eq.~(\ref{separablepotential}) corresponds to
the Weinberg-Tomozawa term plus the $s$-channel nucleon-pole
diagram obtained from the Lagrangian of Eq.~(\ref{lolagr}). Note
that the $u$-channel nucleon-pole diagram cannot be written in the
form of Eq.~(\ref{vmatrix}).}
\begin{equation}
V^{ba}\left( p_2,q_2;p_1,q_1\right)=
-\frac{\epsilon^{bac}\tau^c}{4\,F^2}\,\left( q_1
\hspace{-.85em}/\hspace{.1em}+q_2
\hspace{-.85em}/\hspace{.1em}\right)
-\frac{i\,\stackrel{\circ}{g_A}^2\,\tau^b\tau^a}{4\,F^2}\,\frac{q_2
\hspace{-.85em}/\hspace{.1em}\,\left( p
\hspace{-.45em}/\hspace{.1em}-m\right)\,q_1
\hspace{-.85em}/\hspace{.1em}}{p^2-m^2}\,.\label{separablepotential}
\end{equation}
\noindent In this case the vertex functions $\Gamma^I$ can also be
written as
\begin{equation}
\Gamma^I\left( p_2,q_2;p_1,q_1\right)=\left(%
\begin{array}{cc}
  1 & q_2
\hspace{-.85em}/\hspace{.1em} \\
\end{array}%
\right)\left(%
\begin{array}{cc}
\tau_{11}^I & \tau_{12}^I \\
\tau_{21}^I & \tau_{22}^I \\
\end{array}%
\right)\left(%
\begin{array}{c}
  1 \\
  q_1
\hspace{-.85em}/\hspace{.1em} \\
\end{array}%
\right)\,. \label{tmatrix}
\end{equation}

\begin{figure}
\epsfig{file=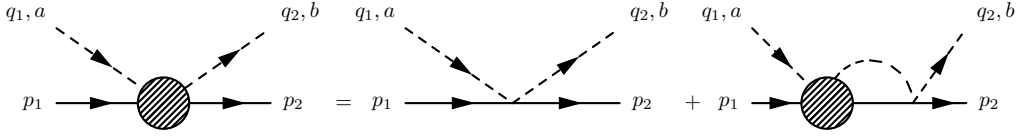, width=14truecm}
\caption[]{\label{piN:fig} Equation for the pion-nucleon
scattering amplitude.}
\end{figure}

\noindent Substituting Eqs.~(\ref{vmatrix}) and (\ref{tmatrix}) in
Eq.~(\ref{decoupledequations}) results in the following matrix
equations
\begin{equation}
\tau^{I} = v^{I} +i\, v^{I}\,g\,\tau^{I}\,,
\label{matrixcoupledequations}
\end{equation}
where
\begin{equation}
g=i \int\frac{d^n k}{(2\,\pi )^n}\, \left(%
\begin{array}{c}
1 \\
k
\hspace{-.45em}/\hspace{.1em} \\
\end{array}%
\right)\,G(p-k,k)\,\left(%
\begin{array}{cc}
1 & k
\hspace{-.45em}/\hspace{.1em} \\
\end{array}%
\right)=\left(%
\begin{array}{cc}
g_{11} & g_{12} \\
g_{21} & g_{22} \\
\end{array}%
\right)\,. \label{gmatrix}
\end{equation}
For the undressed propagator
\begin{equation}
G(p-k,k)=\frac{i}{p \hspace{-.45em}/\hspace{.1em}-k
\hspace{-.45em}/\hspace{.1em}-m+i\,0^+}\,\frac{i}{k^2-M^2+i\,0^+}\,,
\label{undressed prop}
\end{equation}
we obtain
\begin{eqnarray}
g_{11} &=& m\,I_{N \pi}(-p,0)+p
\hspace{-.45em}/\hspace{.1em}\,\left[I_{N \pi}(-p,0)-I_{N \pi
}^{(p)}(-p,0)\right]\,,\nonumber\\
g_{12} &=& g_{21} = \left( p^2+m\,p
\hspace{-.45em}/\hspace{.1em}\right) \,I_{N \pi
}^{(p)}(-p,0)-M^2\,I_{N \pi}(-p,0)-I_N\,,\nonumber\\ g_{22} &=& p
\hspace{-.45em}/\hspace{.1em}\,\left( p^2-m^2\right) \,I_{N \pi
}^{(p)}(-p,0)-M^2\left( p
\hspace{-.45em}/\hspace{.1em}-m\right)\,I_{N \pi}(-p,0)-\left( p
\hspace{-.45em}/\hspace{.1em}-m\right)\,I_N\,.\label{gmatrixelements}
\end{eqnarray}
The loop integrals $I_{N \pi}(-p,0)$, $I_N$, and $I_{N \pi
}^{(p)}(-p,0)$ are given in the appendix.

Decomposing the matrices as
\begin{eqnarray}
v^{I} &=& v_s^{I} + p
\hspace{-.45em}/\hspace{.1em}\, v_v^{I}\,,\nonumber\\
g &=& g_s+p \hspace{-.45em}/\hspace{.1em}\,
g_v\,,\nonumber\\
\tau^{I} &=& \tau_s^{I} + p \hspace{-.45em}/\hspace{.1em}\,
\tau_v^{I}\,, \label{representation}
\end{eqnarray}
and substituting the result in Eq.~(\ref{matrixcoupledequations})
we obtain
\begin{eqnarray}
\tau^{I}_s &=& v^{I}_s +i\, v^{I}_s\,g_s\,\tau^{I}_s+i\,p^2\,
v^{I}_s\,g_v\,\tau^{I}_v +i\,p^2\,
v^{I}_v\,g_s\,\tau^{I}_v+i\,p^2\, v^{I}_v\,g_v\,\tau^{I}_s\,,
\nonumber\\
\tau^{I}_v &=& v^{I}_v +i\, v^{I}_v\,g_s\,\tau^{I}_s+i\,
v^{I}_s\,g_v\,\tau^{I}_s+i\, v^{I}_s\,g_s\,\tau^{I}_v +i\,p^2\,
v^{I}_v\,g_v\,\tau^{I}_v\,.\label{mdecdeqs}
\end{eqnarray}
If we define
\begin{eqnarray}
v^{I}_\pm &=& v_s^{I} \pm \sqrt{p^2}\, v_v^{I}\,,\nonumber\\
g_\pm &=& g_s\pm \sqrt{p^2}\,g_v\,,\nonumber\\
\tau^{I}_\pm &=& \tau_s^{I} \pm \sqrt{p^2}\,\tau_v^{I}\,,
\label{secondrepresentation}
\end{eqnarray}
and substitute in Eq.~(\ref{mdecdeqs}) the resulting equations
reduce to the decoupled system
\begin{equation}
\tau^I_\pm=v^I_\pm + i\,v^I_\pm\,g_\pm\, \tau^I_\pm\,.
\label{redeqs}
\end{equation}

\begin{figure}
\epsfig{file=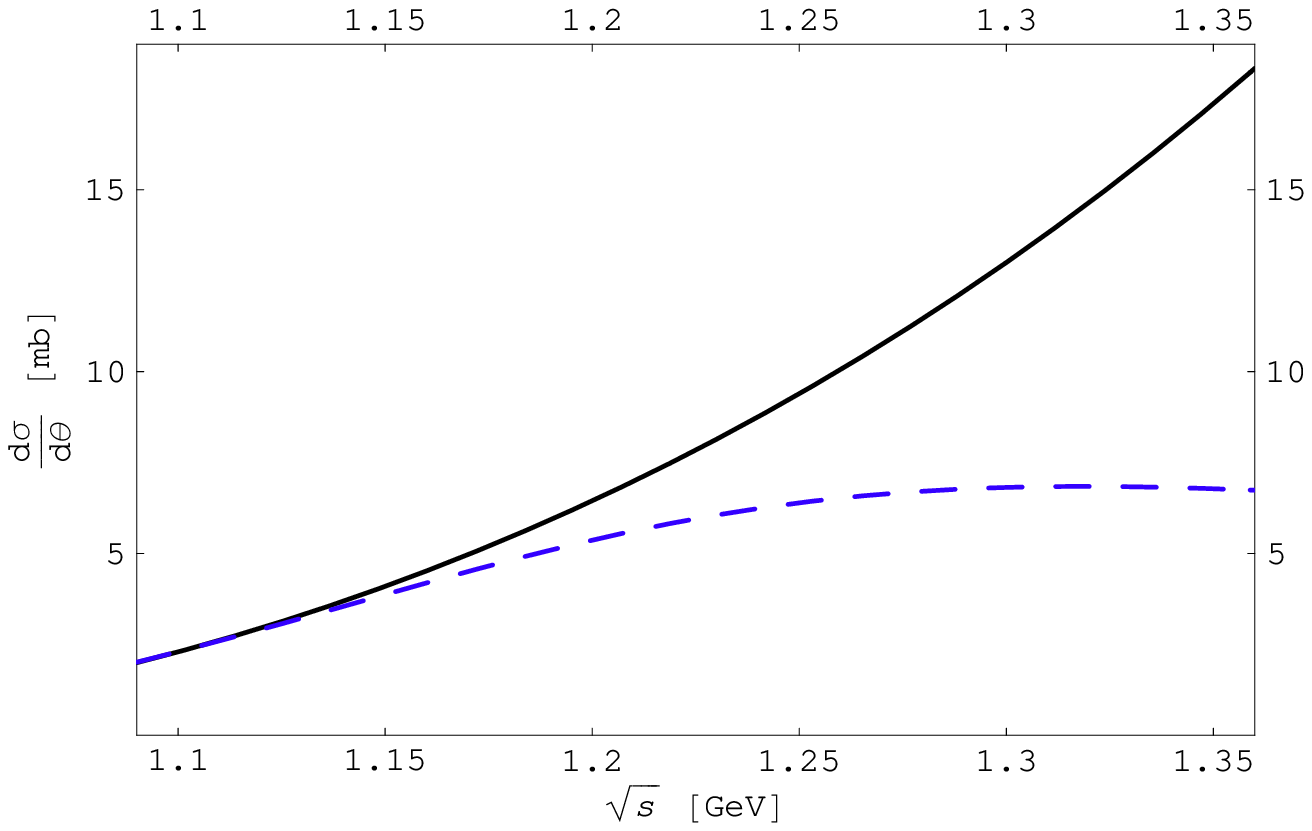, width=10truecm}
\caption[]{\label{piN1forward:fig} Differential cross section for
$\pi^- p\to \pi^0 n$ scattering in forward direction. The solid
and dashed lines correspond to the non-perturbative and
perturbative (tree plus one-loop order) results, respectively. }
\end{figure}

\noindent Eqs.~(\ref{mdecdeqs}) and (\ref{redeqs}) are systems of
matrix equations and can be solved exactly. Inserting the
solutions in
\begin{eqnarray}
\tau^+_s &=& \frac{1}{3}\,\left(2\,\tau^{3/2}_s+\tau^{1/2}_s\right)\,,\nonumber\\
\tau^+_v &=&
\frac{1}{3}\,\left(2\,\tau^{3/2}_v+\tau^{1/2}_v\right)\,,\nonumber\\
\tau^-_s &=& \frac{1}{3}\,\left(\tau^{1/2}_s-\tau^{3/2}_s\right)\,,\nonumber\\
\tau^-_v &=&
\frac{1}{3}\,\left(\tau^{1/2}_v-\tau^{3/2}_v\right)\,,
\label{tauplusminus}
\end{eqnarray}
and using the most general parity-conserving form for the on-shell
$T$ matrix,
\begin{equation}
T^{\pm} = A^\pm + \frac{1}{2}\,\left( q_1
\hspace{-.85em}/\hspace{.1em}+ q_2 \hspace{-.85em}/\hspace{.1em}
\right)\, B^\pm \,,\label{abdef}
\end{equation}
one can calculate the four Lorentz-invariant amplitudes as
\begin{eqnarray}
A^{\pm} &=& \tau^{\pm}_{s;11}+\left( M^2+2\,p_1\cdot
q_1\right)\,\tau^{\pm}_{s;22}+m_N\,\tau^{\pm}_{v;11} + 2
\,\left( M^2+2\,p_1\cdot q_1 \right)\,\tau^{\pm}_{v;12}\nonumber\\
& &-m_N\,\left(M^2+2\,p_1\cdot q_1\right)\,\tau^{\pm}_{v;22}\,, \nonumber\\
B^{\pm} &=& 2\,
\tau^{\pm}_{s;12}-2\,m_N\,\tau^{\pm}_{s;22}+\tau^{\pm}_{v;11}-2\,
m_N\, \tau_{v;12}^{\pm} + \left( 2\, m_N^2+M^2+2\,p_1\cdot
q_1\right)\, \tau^{\pm}_{v;22} \,. \label{t1t2}
\end{eqnarray}
On the other hand by expanding Eqs.~(\ref{t1t2}) perturbatively we
can compare the results of the term-by-term loop expansion with
the non-perturbative expression and estimate the error of the
perturbative approximation for various kinematics.

\begin{figure}
\epsfig{file=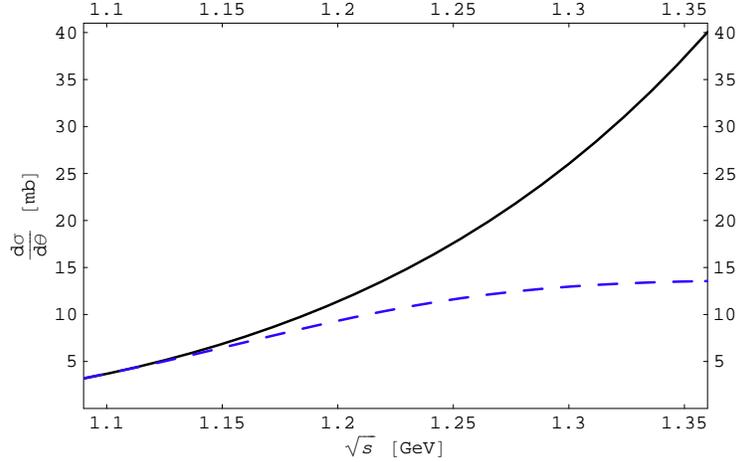, width=10truecm}
\caption[]{\label{piN2forward:fig} The sum of the differential
cross sections for the processes $\pi^- p\to \pi^- p$ and  $\pi^-
p\to \pi^0 n$ in forward direction. The solid and dashed lines
correspond to the non-perturbative and perturbative (tree plus
one-loop order) results, respectively.}
\end{figure}

\begin{figure}
\epsfig{file=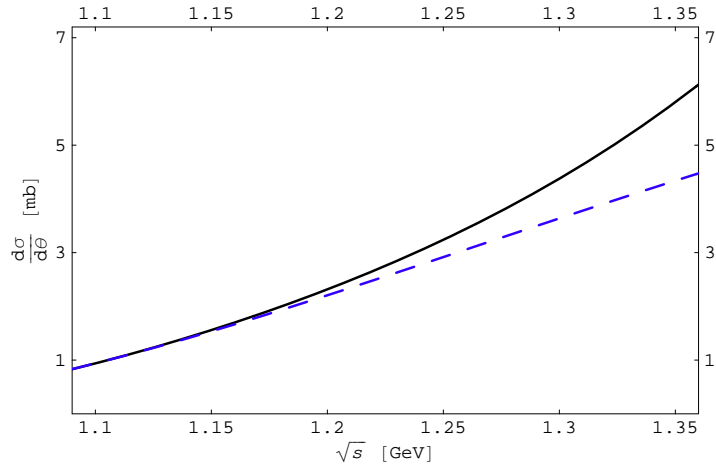, width=10truecm}
\caption[]{\label{piN3forward:fig} Differential cross section for
$\pi^+ p\to \pi^+ p$ scattering in forward direction. The solid
and dashed lines correspond to the non-perturbative and
perturbative (tree plus one-loop order) results, respectively.}
\end{figure}

We calculated exactly (as closed expressions of one-loop
integrals) the non-perturbative (re-summed) contribution and the
tree plus one-loop order contributions to the $\pi^- p\to \pi^0
n$, $\pi^- p\to \pi^- p$, and $\pi^+ p\to \pi^+ p$ scattering
processes for the potential due to the Weinberg-Tomozawa term,
\begin{equation}
V^{ba}\left( p_2,q_2;p_1,q_1\right)=
-\frac{\epsilon^{bac}\tau^c}{4\,F^2}\,\left( q_1
\hspace{-.85em}/\hspace{.1em} +q_2
\hspace{-.85em}/\hspace{.1em}\right) \,.\label{separable0}
\end{equation}
\noindent The results for differential cross sections are given in
Figs.~\ref{piN1forward:fig}-\ref{piN3forward:fig}. These figures
suggest that the perturbative results (tree plus one-loop order)
approximate the re-summed contributions very poorly already for
$s=p^2 \sim m_\Delta^2$.

\section{Nucleon self-energy}

The full (dressed) nucleon propagator has the form
\begin{equation}
i\,S(p)= \frac{i}{p\hspace{-.45em}/\hspace{.1em}-m
-\Sigma(p\hspace{-.45em}/\hspace{.1em})}\,,
\end{equation}
where the nucleon self-energy
$-i\Sigma(p\hspace{-.45em}/\hspace{.1em})$ represents the
one-particle-irreducible contribution to the two-point function.
The nucleon self-energy contains the contributions of
counter-terms so that $m$ corresponds to the nucleon pole mass in
the chiral limit.

\begin{figure}
\epsfig{file=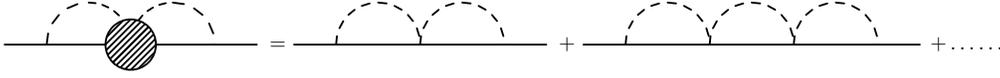, width=14truecm}
\caption[]{\label{Nse:figdiagrams}  Contribution to the nucleon
self-energy.}
\end{figure}

    The physical mass $m_N$ of the nucleon is defined as the solution
to the equation
\begin{equation}
S^{-1}\left( m_N\right)=m_N-m-\Sigma\left( m_N\right)=0\,.
\label{phmass}
\end{equation}
     Below we calculate the contribution to the nucleon mass of an
infinite number of diagrams shown in Fig.~\ref{Nse:figdiagrams}.
The sum of these diagrams can be written in a closed form as
\begin{eqnarray}
-i\,\Sigma &=  & -\frac{\stackrel{\circ}{g_A}^2}{4 \,F^2}\,\int
\int \frac{d^nq_1}{(2\,\pi)^n}\,
\frac{d^nq_2}{(2\,\pi)^n}\nonumber\\
&&\times \, \gamma^5\,q_2 \hspace{-.85em}/\hspace{.1em}\
\frac{1}{p \hspace{-.45em}/\hspace{.1em}-q_2
\hspace{-.85em}/\hspace{.1em}-m_N}\ \tau^b\  \Gamma^{ba}\left(
p_2-q_2,q_2;p_1-q_1,q_1\right)\ \tau^a\ \frac{1}{p
\hspace{-.45em}/\hspace{.1em}-q_1
\hspace{-.85em}/\hspace{.1em}-m_N}\ \gamma^5 q_1
\hspace{-.85em}/\hspace{.1em}\ \nonumber\\
&&\times \, \frac{1}{\left[q_1^2-M^2+i\,0^+\right]\left[
q_2^2-M^2+i\,0^+\right]}\,, \label{NPse}
\end{eqnarray}
where the $\pi N$ vertex function $\Gamma^{ba}$ is obtained by
solving Eq.~(\ref{Tamplequation}) with the potential
\begin{equation}
V^{ba}\left( p_2,q_2;p_1,q_1\right)=
-\frac{\epsilon^{bac}\tau^c}{4\,F^2}\,\left( q_1
\hspace{-.85em}/\hspace{.1em}+q_2
\hspace{-.85em}/\hspace{.1em}\right)\,.\label{seppot}
\end{equation}

\noindent It is easily verified that $\Sigma$ depends only on
$\Gamma^{1/2}$. Using the solution for $\Gamma^{1/2}$ from the
previous section in Eq.~(\ref{NPse}) and integrating over $q_1$
and $q_2$ we obtain an analytic expression for the contribution of
the diagrams in Fig.~\ref{Nse:figdiagrams} to the nucleon mass
\begin{equation}
\delta m = -\frac{3\,\stackrel{\circ}{g_A}^2}{4 \,F^2}\,
\frac{N}{D}\,, \label{contributioninmass}
\end{equation}
where
\begin{eqnarray}
N &=&  \left(m+m_N\right) \left( 4\, F^2-I_\pi\right) \biggl[
\left( m_N-m\right) I_\pi - \left(m^2-M^2-2\,m\,m_N +m_N^2\right)
\left(m+m_N\right) I_{N \pi}\biggr]^2\,, \nonumber\\
D &=& 2 \,m_N \Biggl\{ 8 \,m_N\,F^4+4\,\biggl[ \left(
m-\,m_N\right) I_{\pi} +\left(m^2-M^2-2\,m\,m_N +m_N^2\right)
(m+m_N) \,I_{N \pi} \biggr] F^2\nonumber\\
& & + \, I_{\pi} \biggl[\left( m_N-m\right) I_{\pi} -
\left(m^2-M^2-2\,m\,m_N +m_N^2\right) \left( m+m_N\right) I_{N
\pi}\biggr]\Biggr\}\,. \label{massD}
\end{eqnarray}
On the other hand, by expanding Eq.~(\ref{contributioninmass}) in
powers of $1/F^2$ we can identify the contributions of each
diagram separately. Using the IR renormalization scheme and
substituting $m=882.8$ MeV \cite{Fuchs:2003kq}, $m_N=938.3$ MeV,
$F=92.4$ MeV, $\stackrel{\circ}{g_A}=1.267$ and $M=139.6$ MeV we
obtain
\begin{equation}
\delta m = -0.00233530\,{\rm MeV}=\left( -0.00230219-0.00003305 -
0.00000007 +\cdots\right)\,{\rm MeV}\,. \label{dmBL}
\end{equation}
As can be seen from Eq.~(\ref{dmBL}) the first term in the
perturbative expansion reproduces the non-perturbative result
 well and the higher-order corrections are clearly
suppressed.

\begin{figure}
\epsfig{file=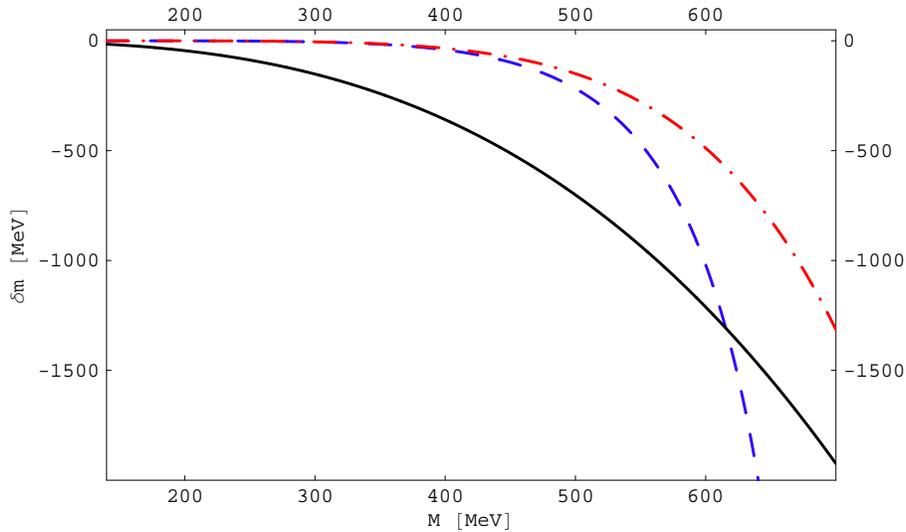, width=12truecm}
\caption[]{\label{Nseplot:fig} Contributions to the nucleon mass
as functions of $M$. The solid, dashed, and dashed-dotted lines
correspond to $\delta m_3$, $\delta m$, and the contribution of
the two-loop diagram in Fig.~\ref{Nse:figdiagrams}, respectively.}
\end{figure}

It is relevant for the chiral extrapolation of lattice data to
consider the nucleon mass for larger values of the quark masses.
Therefore in Fig.~\ref{Nseplot:fig} we plot $\delta m$ of
Eq.~(\ref{massD}) together with the leading contribution (first
diagram in Fig.~\ref{Nse:figdiagrams}) and the leading
non-analytic correction to the nucleon mass $\delta
m_3=-3\,g_A^2\,M^3/(32\,\pi\,F^2)$ \cite{Gasser:1988rb} as
functions of $M$. As can be seen from this figure, up to $M\sim
500$ MeV the non-perturbative sum of higher-order corrections is
suppressed in comparison with the $\delta m_3$ term. Also, the
leading higher order contribution reproduces the non-perturbative
result quite well. On the other hand for $M\gtrsim 600$ MeV the
higher-order contributions are no longer suppressed in comparison
with $\delta m_3$.

\section{Summary and discussion}
In this work we have addressed the issue of convergence of
perturbative calculations in BChPT by analyzing pion-nucleon
scattering and the nucleon mass. By solving the equation for the
$\pi N$ vertex function using dimensional regularization we have
obtained an exact expression for a sum of an infinite number of
loop diagrams. The solution is given in a closed form in terms of
one-loop integrals. We have renormalized the obtained
non-perturbative expression by applying infrared renormalization
\cite{Becher:1999he}. We compared the perturbative contributions
with the re-summed expression for elastic $\pi N$ scattering. We
find that already for $s\sim m_\Delta^2$ the perturbative results
approximate the re-summed non-perturbative expression very poorly.
As the potential, which has been iterated by solving the equation
does not receive contributions from the intermediate $\Delta$
state, we conclude that the inclusion of the $\Delta$ as an
explicit degree of freedom does not solve the problem of
convergence of the considered loop contributions. In our opinion,
to solve this problem one needs to include the $\Delta$ degrees of
freedom \cite{Hacker:2005fh} and {\it simultaneously} consider the
$\pi N$ scattering equations.

Next, using the non-perturbative result for the $\pi N$ scattering
amplitude we have obtained an exact expression corresponding to a
sum of an infinite number of diagrams contributing to the nucleon
self-energy. Using the infrared renormalization scheme and
comparing the non-perturbative contribution to the nucleon mass
with contributions of the first several terms in its perturbative
expansion we conclude that the so obtained perturbative series for
the nucleon mass converges very well. We also considered the
correction to the nucleon mass for larger values of quark masses
and found that the re-summed higher order contributions become
larger than the leading non-analytic contribution for $M\gtrsim
600$ MeV. From this we conclude that for such values of $M$ BChPT
cannot be trusted in extrapolations of lattice data. Even if there
are large cancelations these cannot be treated systematically in
standard BChPT. As we have summed up only a subset of higher-order
diagrams, our analysis is not complete and therefore our result
should be considered as an estimate for an upper limit of the
radius of convergence.

\acknowledgments

We would like to thank D.~Leinweber for useful comments on the
manuscript.
   The work of D.~D. and J.~G. was supported by the Deutsche
Forschungsgemeinschaft (SFB 443).

\section{appendix}

One-loop integrals:
\begin{eqnarray}
\label{Imunpi} I^\mu_{N\pi}(-p,0)&=& p^\mu \,
I^{(p)}_{N\pi}(-p,0)= i\int\frac{d^n k}{(2\pi)^n}
\frac{k^\mu}{[(k-p)^2-m^2+i0^+] [k^2-M^2+i0^+]}\nonumber\\
&=&\frac{p^\mu}{2p^2}\left[(p^2-m^2+M^2)\,I_{N\pi}(-p,0)+I_N
-I_\pi\right]\,,\nonumber\\
I_\pi &=& i\,\int \frac{d^n k}{(2\pi)^n}\frac{1}{k^2-M^2+i0^+}
=2\,M^2\bar{\lambda}+\frac{M^2}{8\pi^2}\ln\left(\frac{M}{m}\right)\,,
\nonumber\\
I_N &=& i\int \frac{d^n k}{(2\pi)^n} \frac{1}{k^2-m^2+i0^+}
=2 \,m^2\,\bar{\lambda}\,,\nonumber\\
\label{INpi} I_{N\pi}(-p,0)&=&i\int\frac{d^n
k}{(2\pi)^n}\frac{1}{[(k-p)^2-m^2+i0^+]
[k^2-M^2+i0^+]}\nonumber\\
&=&2\,\bar{\lambda}+\frac{1}{16\pi^2}\left[-1
+\frac{p^2-m^2+M^2}{p^2}\ln\left(\frac{M}{m}\right)
+\frac{2mM}{p^2}F(\Omega)\right],
\end{eqnarray}
where
\begin{eqnarray*}
\label{lambdabar} \bar\lambda &=& {m^{n-4}\over (4\pi )^2}\left\{
{1\over n-4}- {1\over 2}\left[ \ln (4\pi )+\Gamma
'(1)+1\right]\right\}\,,\nonumber\\
 F(\Omega) &=&
\left \{
\begin{array}{ll}
\sqrt{1-\Omega^2}\arccos(-\Omega),&-1\leq\Omega\leq 1,\\
\sqrt{\Omega^2-1}\ln\left(\Omega+\sqrt{\Omega^2-1}\right)
-i\pi\sqrt{\Omega^2-1},&1\leq \Omega,
\end{array} \right.\,,\nonumber\\
\Omega &=&\frac{p^2-m^2-M^2}{2\,m\,M}\,.
\end{eqnarray*}
Infrared renormalized expression for $I_{N\pi}(-p,0)$:
\begin{eqnarray}
\label{INpiBL} I^{IR}_{N\pi}(-p,0)&=& 
\frac{1}{16\pi^2}\,\frac{p^2-m^2+M^2}{2\,p^2}\,\left[ 2\,
\ln\left(\frac{M}{m}\right)-1\right]+F_{IR}(\Omega),
\end{eqnarray}
where
\begin{eqnarray*}
F_{IR}(\Omega) &=& \left \{ \begin{array}{ll}
\frac{1}{8\pi^2}\,\frac{m\,M}{p^2}\,\sqrt{1-\Omega^2}
\,\arccos\left(-\frac{\alpha+\Omega}{\sqrt{1+2\,\alpha\,\Omega+\alpha^2}}\right),&-1\leq\Omega\leq 1,\\
-\frac{1}{16\, \pi ^2}\ \frac{m\,M}{p^2}\, \sqrt{\Omega^2-1} \ \ln
\,\frac{\alpha +\Omega -\sqrt{\Omega ^2-1}}{\alpha +\Omega
+\sqrt{\Omega ^2-1}}-\frac{i}{8\, \pi}\ \frac{ m\,M }{p^2}\,
\sqrt{\Omega^2-1}\,,&1\leq \Omega,
\end{array} \right.
\end{eqnarray*}
and $\alpha=\frac{M}{m}$\,.

\begin{equation}
I_{N\pi}=I_{N\pi}(-p,0)|_{p^2=m_N^2}\,. \label{Inpi}
\end{equation}

\end{document}